\documentclass[12pt,fleqn]{article}
\usepackage{psfig,float}
\usepackage{graphicx}
\usepackage{bm}

\begin{document}

\begin{center}
{\Large{\bf The $(\pi^-,\gamma\gamma)$ reaction
in nuclei and the $\sigma$ meson in the medium}}
\end{center}
\vspace{1cm}

\begin{center}
{\large{L. Roca and E. Oset}}

 \vspace{.5cm}
{\it Departamento de F\'{\i}sica Te\'orica and IFIC,
 Centro Mixto Universidad de Valencia-CSIC\\
 Institutos de Investigaci\'on de Paterna, Apdo. correos 22085,
 46071, Valencia, Spain}    

\end{center}

\begin{abstract}
A theoretical analysis of the $(\pi^-,\gamma\gamma)$ reaction in 
nuclei is made in order to 
find the viability of this reaction to test modifications
of the $\sigma$ meson mass in nuclear matter.
The $\pi\pi$
correlation in the scalar-isoscalar channel in nuclear matter
could, in principle, manifest itself
in this reaction since it plays an important
role in the $\pi\pi\to\gamma\gamma$ mechanism. But we conclude
that this effect is hardly visible in this reaction due to the
strong background of the pion-Bremsstrahlung terms. 
Only with some special cuts and for some polarization states are the effect
visible at the cost of a strong reduction in the cross section. 

\end{abstract}


\section{Introduction}

The study of the $\pi^-$-induced two photon emission on nuclei has attracted much attention in
the last decades. But most of the efforts have been aimed at its application in pionic atoms
\cite{Gil:1994mi,Ericson:1975ha,Barshay:hg,Lapidus,Nyman:1977gk}.
In this work we will aim in another direction,
focusing in two main objectives: First of all we will explain how to extend the free
$\pi^-p\to\gamma\gamma n$ reaction to the process in nuclei using many body techniques
successfully used to describe 
different pionic reactions \cite{Salcedo:md,Oset:qd}, 
 paying special attention to the distortion of the
initial pion in the nucleus. Our second aim will be to study the viability of using the
$(\pi^-,\gamma\gamma)$ reaction in nuclei as a way to test the modification of the properties of
the scalar-isoscalar $\sigma$ meson in nuclear matter. 

In the recent past there has been a very lively discussion about the existence,
nature and properties of the scalar meson $\sigma(500)$ which
lasts till today \cite{kyoto,Beveren:2002gy}.
The most important source of controversy comes from the confrontation between the interpretation
of the $\sigma$ meson as an ordinary $q\overline{q}$ meson or as a $\pi\pi$ resonance. 
The advent of $\chi PT$ has
brought new light into this problem and soon it was suggested 
\cite{Gasser:1991bv,Meissner:1991kz} that the $\sigma$
could not qualify as a genuine meson which would survive in the limit of large
$N_c$. The reason is that
the  $\pi \pi$ interaction in s-wave in the isoscalar sector is strong enough
to generate a resonance through multiple 
scattering of the pions. This seems to be the case, and even in models starting
with a seed of $q \bar{q}$ states, the incorporation of the $\pi \pi$
channels in a unitary approach leads to a large dressing by a pion
cloud which makes negligible the effects of the original $q \bar{q}$ seed 
\cite{torn}.  This idea has been made more quantitative through the introduction
of the unitary extensions of $\chi PT$ ($U \chi PT$) 
\cite{Dobado:1990qm,Dobado:1993ha,OllOsePel,Oller:1999zr}. These works implement
unitarity in coupled channels in an exact form and use the input of the lowest
and second order chiral Lagrangians of \cite{Gasser:1985ux}.

Another point of interest which can help us understand the nature of the
 $\sigma$ meson is the modification of its properties at finite nuclear density. 
The importance of the medium modification of the $\pi \pi$ interaction in the scalar
sector was suggested in \cite{Schuck:1988jn} where the $\pi \pi$ amplitude in
the medium developed large peaks below the two pion threshold, somehow
indicating that the $\sigma$ pole had moved to much lower energies.  The issue
has been revised and the models have been polished incorporating chiral
constraints \cite{Rapp:1996ir,Aouissat:1995sx,Chiang:1998di} with the result 
that the peaks
disappear at normal density, 
but still much strength is shifted to low energies.

Experimental tests of the renormalization of the $\sigma$ properties in the
medium have been performed using two pion production reactions induced by
photons $(\gamma,\pi\pi)$ \cite{Metag} and $(\pi,\pi\pi)$ 
\cite{Bonutti:1998zw,Starostin:2000cb,Starostin:2002rv}.
In the latter reaction the absorption of the pions
in the nucleus caused the process to be too much peripheral to manifest
the in-medium $\sigma$ effects,
to the point that the large changes seen in the experimental $\pi\pi$ mass distribution
from deuterium to heavier nuclei \cite{Bonutti:1998zw,Starostin:2000cb}
could not be explained \cite{VicenteVacas:1999xx}. It was argued there that the abnormal
feature was not the size of the $\pi^+\pi^-$ invariant mass distribution close to threshold
in nuclei, but the very small size of this magnitude in deuterium,
coming from a subtle
cancellation of different terms in the amplitude. The offset of this cancellation in nuclei
could bring the strength of the $\pi^+\pi^-$ distribution in nuclei to its "normal
size" (given by the $\pi^+\pi^+$ distribution).

So far the most successful reaction has been the $(\gamma,\pi^0\pi^0)$ reaction where the
$\pi^0\pi^0$ invariant mass distribution in nuclei, evaluated in \cite{Roca:2002vd}, shows
a shift of strength towards lower invariant masses with respect to the reaction on the
proton. This shift has been corroborated in a recent experiment in \cite{Metag}.

Because of the importance and controversy on this topic, more reactions testing the
modification of the $\sigma$ meson in the medium would be welcome.
 One of the possible candidates is the
$(\pi^-,\gamma\gamma)$ reaction in nuclei. This process is interesting because
 there is no absorption
of the final state particles by the nucleus 
and, hence, allows one to test bigger nuclear densities.
In this sense, the reaction should in principle be preferable to other reactions which have
been theoretically used as a test of the modification of the $\pi\pi$ interaction in the
nuclear medium like $(\gamma,\pi\pi)$ and $(\pi,\pi\pi)$ in nuclei. 

The use of the $(\pi^-,\gamma\gamma)$ reaction to test experimentally the in-medium
$\sigma\to\gamma\gamma$ modification has been preliminary study in \cite{Starostin:2002rv}.
In this latter work, comparison of data on $(\pi^-,\gamma\gamma)$ for proton and $^{12}C$
targets, using simulations of the theoretical model of \cite{Chiku:1998kd}, hinted, as a
preliminary result, to a small modification of the 
$\sigma\to\gamma\gamma$ decay in the nuclear medium. 

From the point of view of $U\chi PT$
 the $\sigma$ meson can appear in the
$(\pi^-,\gamma\gamma)$ reaction via the $\pi\pi$ rescattering involved in the
$\pi^+\pi^-\to\gamma\gamma$ mechanism \cite{Oller:1997yg}. But this mechanism
is only one among all the mechanisms involved in the $(\pi^-,\gamma\gamma)$ reaction
\cite{Cammarata:sj}. The strength of the terms in the reaction proceeding through $\sigma$
excitation, relative to that of other mechanisms, will determine the chances to see
$\sigma$ effects in the medium. Hence a quantitative study of this reaction is needed to
provide a proper answer to this question.


\section{Free reaction}

\begin{figure}[h]
\centerline{\hbox{\psfig{file=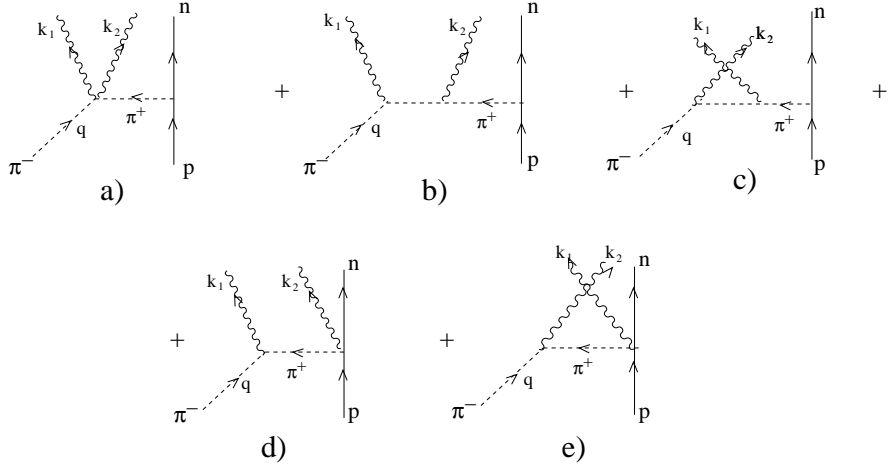,width=12cm}}}
\caption{\rm 
Tree level diagrams for the $\pi^-p\to \gamma\gamma p$ reaction}
\label{fig:diagrams_free}
\end{figure}

For the evaluation of the $\pi^-p\to\gamma\gamma n$ amplitude we will
first consider the mechanisms shown in Fig.~\ref{fig:diagrams_free}.

The $d$ and $e$ diagrams, which are
negligible at small momenta, were not considered in \cite{Gil:1994mi}
since the authors were only interested in pionic atoms. But now, since
we are also interested in higher momenta of the initial pion, one must keep these terms.
Other possible diagrams like those where the two photons come from one 
$\pi^0$ decay or the Bremsstrahlung of one photon on the nucleon lines
were considered in \cite{Cammarata:sj}.
But we will not consider them because: In the first case the diagrams
with the $\pi^0\to\gamma\gamma$ mechanism are only relevant near the region where
$M_{\gamma\gamma}\sim m_{\pi}$ and can be
easily filtered experimentally. 
In the second case, the diagrams with
Bremsstrahlung on the nucleon lines can be neglected because their contribution
to the cross section are of
the order of $\mathcal{O}(p/2M)$, with $p$ the momenta of the nucleons and the photon involved
and $M$ the mass of the nucleon.
It is also possible to consider the effect of intermediate $\Delta(1232)$ states 
which in \cite{Beder:fr} were estimated to be 
around $7\,$\% of the diagram $a$ of Fig.~\ref{fig:diagrams_free} near threshold.
The smallness of these terms was also corroborated by the fair agreement of the model
ignoring them with the data of $(\pi^-,\gamma\gamma)$ in pionic atoms \cite{Gil:1994mi}.
At the higher energies where we will work, the $\Delta$ contribution
should be
a little bit bigger. Yet, for the purpose of the present paper, which is to build up  the
framework for the study of the $(\pi,\gamma\gamma)$ reaction in nuclei
and its viability to see
$\sigma$ medium modification, we can safely ignore the nucleon-Bremsstrahlung and
$\Delta$ terms.

The amplitude corresponding to the diagrams of Fig.~\ref{fig:diagrams_free},
 in the Coulomb gauge, takes the form

\begin{eqnarray}\nonumber
T&=&-i2\sqrt{2}e^2\frac{f}{m_{\pi}}\vec{\sigma}\cdot\vec{q}\,'
\, \frac{1}{q'^2-m_{\pi}^2}  \left[\vec{\epsilon}_1\cdot\vec{\epsilon}_1
+\frac{\vec{q}\cdot\vec{\epsilon_1}\,\vec{q}\,'\cdot\vec{\epsilon_2}}{q\cdot k_1}
+\frac{\vec{q}\cdot\vec{\epsilon_2}\,\vec{q}\,'\cdot\vec{\epsilon_1}}{q\cdot k_2}
\right]-\\
&&-i\sqrt{2}e^2\frac{f}{m_{\pi}}
\left[\frac{\vec{q}\cdot\vec{\epsilon_1}\,\vec{\sigma}\cdot\vec{\epsilon_2}}{q\cdot k_1}
+\frac{\vec{q}\cdot\vec{\epsilon_2}\,\vec{\sigma}\cdot\vec{\epsilon_1}}{q\cdot k_2}
\right]
\label{eq:amplitude_Born}
\end{eqnarray}
The terms represent the diagrams $a$, $b$, $c$, $d$ and $e$
of Fig.~\ref{fig:diagrams_free}
respectively,
$\vec{\epsilon_1}$ and $\vec{\epsilon_2}$ are the polarization vectors of the two photons,
$q'=k_1+k_2-q$ and $f\approx 1$. 
The first three terms relate to the Born terms in $\gamma\gamma\to\pi\pi$ and
we shall call them also Born terms in
the $(\pi^-,\gamma\gamma)$ reaction.\\

The cross section for the $\pi^-p\to\gamma\gamma n$ reaction is given by

\begin{eqnarray}\label{total}
\sigma &=& \frac{M}{\lambda^{1/2}(s, m_{\pi}^2, M^{2})}
\frac{1}{2(2\pi)^{5}}
\int \frac{d^{3} k_{1}}{2\omega_{1}}
\int \frac{d^{3} k_{2}}{2\omega_{2}}
\int d^{3}p_{2}\frac{M}{E_{2}} \cdot \\ 
&&\cdot \delta^{4} (q + p_{1} - p_{2} - k_{1} -k_{2})
\overline{\sum_{s_i}} \sum_{s_f} |T|^{2} \nonumber \\ \nonumber
&=&\frac{M^{2}}{\lambda^{1/2}(s, m_{\pi}^2, M^{2})}
\, \frac{1}{8(2\pi)^{4}}
\int d \omega_{2} d \omega_{1} d \cos \theta_{2} d \phi_{12}
\ \theta (1 - cos^{2} \theta_{12}) \overline{\sum_{s_i}} \sum_{s_f,\lambda} |T|^{2}
\end{eqnarray}
where $q = (\omega,\vec{q}\,)$, $p_{1} = (E_{1},\vec{p}_{1})$,
$p_{2} =(E_{2},\vec{p}_{2})$,
$k_{1} = (\omega_{1}, \vec{k}_{1})$, $k_{2} =
(\omega_{2},\vec{k}_{2})$
are the momenta
of the pion, initial proton, outgoing neutron
and the outgoing photons respectively. The label $\lambda$ indicates
that the sum is done over all the polarizations of the photons.
In Eq.~(\ref{total}) 
$\phi_{12}$,
$\theta_{12}$ are the azimuthal and polar angles of
$\vec{k}_{1}$ with
respect to $\vec{k}_{2}$ and $\theta_{2}$ is the angle of
$\vec{k_{2}}$ with
the $z$ direction defined by the incident pion momentum
$\vec{q}$. While $\phi_{12}$ 
is an integration variable, $\theta_{12}$ is given by
 energy-momentum conservation in terms
of the other variables. $T$ is the
invariant matrix element for the reaction.

The pion-Bremsstrahlung mechanisms, $b$, $c$, $d$ and $e$ diagrams, 
have a typical infrared divergence when the momentum of the pion is not zero,
thus some cut in the photon energy is needed.

\begin{figure}[h]
\centerline{\hbox{\psfig{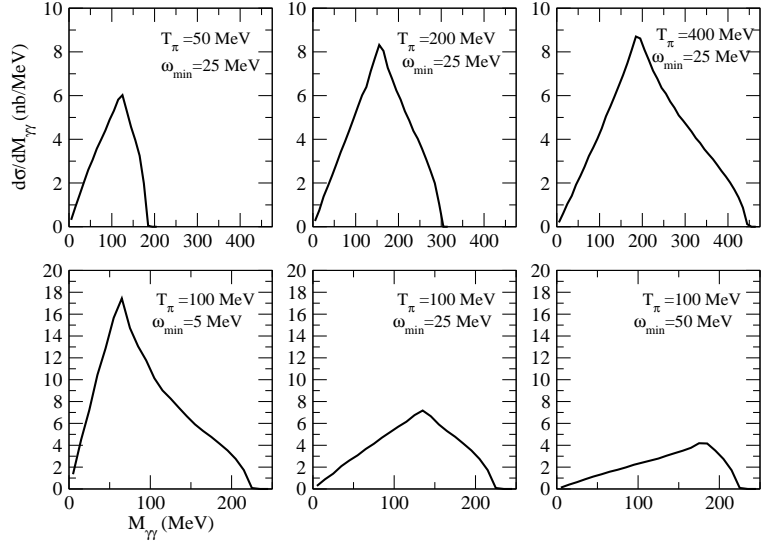}}}
\caption{\rm 
Invariant mass distribution of the two photons: Upper row: varying the pion kinetic energy
$(T_{\pi})$ at a fixed cut in the photon energy $(\omega_{min})$.
Lower row: varying $\omega_{min}$ at a fixed $T_{\pi}$.}
\label{fig:z1}
\end{figure}

 In the upper row of Fig.~\ref{fig:z1} we can see the 
invariant mass distributions for the
two photons at different kinetic energies of the pion, calculated
removing those events where the energy of some of the photons is less than $25\,MeV$.

\vspace{0.9cm}

In the lower row of Fig.~\ref{fig:z1} we can also see the invariant mass
distributions
of the two photons at a fixed
kinetic energy of the pion ($100\,MeV$) but varying the cut in the photon energy.
We note the trend of the cross section blowing up as the energy of the photons
goes to zero, because of the infrared divergence of the
pion-Bremsstrahlung terms. Both the shape and strength of the invariant mass distribution
depend crucially on the cut in the energy of the photons.

\section{The $(\pi^-,\gamma\gamma)$ reaction in nuclei}

The cross section for the process in nuclei can be calculated
using the accurate and simple many body techniques 
summarized in \cite{Carrasco:vq}, successfully used in
many pion and lepton interactions with
nuclei \cite{Roca:2002vd,Chiang:nh,Singh:dc,Oset:qd,Salcedo:md,Gil:1997bm}.

\begin{figure}[h]
\centerline{\hbox{\psfig{file=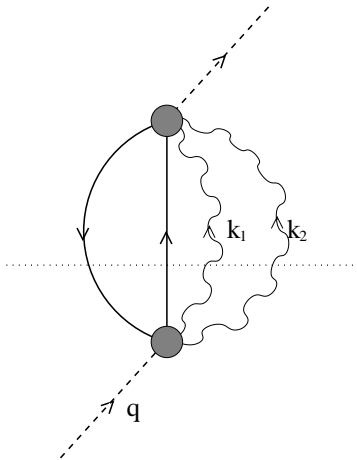,width=5cm}}}
\caption{\rm 
Diagram for the $\pi$ selfenergy having as a source of imaginary part the
cut (dotted line) of the two photon lines and the particle-hole excitation.}
\label{fig:self-pi}
\end{figure}

The total cross section can be related to the imaginary
part of the pion selfenergy of Fig.~\ref{fig:self-pi} through the
expression  

\begin{equation}\label{eq:sigma-a}
\sigma=-\frac{1}{q}\int{d^3\vec{r}\, Im\Pi(q,\rho(r))}
\end{equation}
where $\Pi(q,\rho (r))$ is the pion selfenergy of the diagram
of Fig.\ref{fig:self-pi} and $q$ is the pion momentum.
Equation (\ref{eq:sigma-a}) is making an implicit use of the local density
approximation, since the photon selfenergy is evaluated at the nuclear density
at the point $\vec{r}$ in the integral.

 The pion selfenergy corresponding to the diagram of Fig.~\ref{fig:self-pi}
  is given by
 
 \begin{eqnarray} \label{eq:self}
-i\Pi(q,\rho(r))=\int \frac{d^4k_1}{(2\pi)^4} \int \frac{d^4k_2}{(2\pi)^4}
\,\frac{i}{k_1^2+i\epsilon}
\,\frac{i}{k_2^2+i\epsilon}
\,\frac{1}{2}i\overline{U}(q-k_1-k_2,\rho(r))
\frac{1}{2}\sum_{s_i,s_f,\lambda}
(-i)^2T^2
\end{eqnarray}
where $\overline{U}$ is the Lindhard function which accounts for the
particle-hole excitation

\begin{equation}\label{eq:Lindhard}
\overline{U}(q,\rho)=2\int\,\frac{d^3k}{(2\pi)^3}
\left[\frac{n(k,\rho)[1-n(k-q,\rho)]}{q^0+E(k)-E(k+q)+i\epsilon}
+ \frac{n(k,\rho)[1-n(k+q,\rho)]}{-q^0+E(k)-E(k-q)+i\epsilon}\right]
\end{equation}
and the last $1/2$ factor is the symmetry factor for the two photons.
In this expression $E(p)=\sqrt{|\vec{p}|^2+M^2}$,
$n(p,\rho(\vec{r}))$ is the occupation number of the local Fermi sea,
which is unity for $\mid \vec{p}\mid \leq k_F(\vec{r})$ and zero
for $\mid \vec{p}\mid > k_F(\vec{r})$ and  
$k_F(\vec r)=[\frac{3}{2}\pi^2\rho(\vec r)]^{1/3}$ is the Fermi momentum.
One can see in Eq.~(\ref{eq:Lindhard})
the effect of the Fermi motion of the initial nucleon, $d^3k\,n(k,\rho)$,
and the Pauli blocking of the final nucleon, $1-n(k+q,\rho)$.

The imaginary part of the pion selfenergy is obtained when the
intermediate states (a particle-hole and two photons) are placed on shell in the
integrations over the momenta of the intermediate states.
This can actually be implemented using Cutkosky rules, applying the following 
substitutions to the lines cut by a straight line drawn between the two
external pions as shown in Fig.~\ref{fig:self-pi}

\begin{eqnarray}
\nonumber \Pi(k)&\quad\to\quad& 2i\,Im\Pi(k) \\ \label{eq:cutkosky}
\overline{U}(k)&\quad\to\quad& 2i\,\theta(k^0)Im\overline{U}(k) \\
\nonumber D(k)&\quad\to\quad& 2i\,\theta(k^0)Im D(k)
\end{eqnarray}
with D(k) the photon propagator $\frac{1}{k^2+i\epsilon}$ and conjugating the $T$ matrix
in the upper vertex.\\

Applying these rules to Eq.~(\ref{eq:self}) and using Eq.~(\ref{eq:sigma-a}) we
obtain the following expression for the $(\pi^-,\gamma\gamma)$ reaction in nuclei:
\begin{eqnarray} \label{eq:X-IM}
\sigma=-\frac{1}{16(2\pi)^6}\frac{1}{q}\int d^3r \int d^3k_1 \int d^3k_2
\,\frac{1}{k_1 k_2}
\,Im\overline{U}(q-k_1-k_2,\rho(r))
\sum_{s_i,s_f,\lambda}|T|^2
\end{eqnarray}

So far the distortion of the incoming pion inside the nucleus before it
reaches the production point $\vec{r}$ has not been taken into account.
Since pions interact strongly with nucleons, the loss of pion flux makes the
reactions in nuclei involving pions in initial or final states to be more
peripheral and to produce less desired events. This
strong distortion of the pions inside the nucleus has turned out to be crucial in
the evaluation of cross sections and nuclear effects in reactions
involving pions \cite{VicenteVacas:1999xx,Roca:2002vd}.

Pions inside a nucleus can be distorted in many ways: 
Can be absorbed; can undergo quasielastic collisions and change direction, energy, charge or
even have inelastic collisions and produce more pions.
In order to quantify the distortion of the
pions we make an eikonal approximation and remove from the pion flux those
pions which undergo absorption, which indeed disappear. We also remove the pions which
undergo quasielastic collisions because, even if they do not disappear,
they lose much energy
and the cross section for production of two photons in the upper part of the $2\gamma$
spectrum, where we will be interested, is considerably reduced, (see Fig.~\ref{fig:z1}).
Thus we have to include in the integrand of Eq.~(\ref{eq:X-IM}) an 
eikonal factor for this initial state interaction (ISI) of the pion which is given by 

\begin{eqnarray}
\hspace{-1.1cm}
F_{ISI}(\vec{r},\vec{q})&=&exp\left[-\int_{-\infty}^{0}dl
\,\mathcal{P}(\vec{r}\,',\vec{q})
\right]
\label{eq:eikonal}
\\ \nonumber
\\ \nonumber  \vspace{0.3cm}
& &\hspace{-2cm}
 \vec{q}:\textrm{momentum of the pion}
\\ \nonumber \vspace{0.3cm}
& &\hspace{-2cm}
 \vec{r}:\textrm{production point inside the nucleus}
\\ \nonumber  \vspace{0.3cm}
& &\hspace{-2cm}
\vec{r}\,'=\vec{r}+l\ \vec{q}/\mid \vec{q}\mid:\textrm{ integration point
 in the $\pi^-$ trajectory}
\\ \nonumber  \vspace{-0.5cm}
& &\hspace{-2cm}
\mathcal{P}(\vec{r}\,',\vec{q}): \textrm{ reaction probability per unit length
of the $\pi^-$ in the nucleus }
\end{eqnarray}
The interpretation of the eikonal factor (\ref{eq:eikonal}) is very intuitive
since it is nothing but an exponential decay law which represents the 
probability for a $\pi^-$ to reach the production point $\vec{r}$ without interacting with
the nucleus. The probability of absorption or quasielastic collisions per unit length depends
crucially on the energy of the pion, and there are accurate models for different energy regions to
account for it: From zero to $\sim 300\,$MeV we will use 

\begin{equation}
\mathcal{P}(\vec{r}\,',\vec{q})=-\frac{1}{\mid \vec{q}\mid} Im \Pi(\vec{r}\,',\vec{q})
\end{equation}
where $\Pi$
is the pion selfenergy in the nuclear medium. For the region 
of very low energy pions (from $0$ to
$\sim 85\,$MeV) we will use for $\Pi$ the model of \cite{Nieves:ye},
based on a extrapolation for low energy pions of the
pion-nucleus optical potential developed for pionic atoms using many body
techniques. In this work the
 imaginary part of the potential is split into a part that
accounts for the probability of quasielastic collisions and another one which
accounts for the pion absorption probability. In the $\Delta$ resonance region (from $\sim 85$ to
$\sim 300\,$MeV) we use for $\Pi$  the model of \cite{Salcedo:md,Oset:qd}
which considers absorption by two
and three body mechanisms and quasielastic collisions at the level of $1p1h$ and $2p2h$
excitation. For kinetic energies of the pion beyond $\sim 300\,$MeV we will use for the
probability of distortion per unit length
 
\begin{eqnarray}\hspace{-1.1cm}\label{eq:eikonal2}
\mathcal{P}(\vec{r}\,',\vec{q})&=&\sigma_{\pi^- p}
\rho_p(\vec{r}\,')+\sigma_{\pi^- n}\rho_n(\vec{r}\,')+C^{abs.(2)}\rho^{2}(\vec{r}\,')
+C^{abs.(3)}\rho^{3}(\vec{r}\,')
\\ \nonumber
& & \hspace{-1cm}
\sigma_{\pi^-N}:\textrm{quasielastic $\pi^-$-Nucleon cross section}
\\ \nonumber \vspace{-0.5cm}
& &\hspace{-1cm}
C^{abs.}:\textrm{absorption coefficients by 2 and 3 body mechanisms}
\end{eqnarray}
where $\sigma_{\pi^-N}$ are taken from experiment while $C^{abs}$ are calculated theoretically
from the model of \cite{Oset:1990zj}. The superindex $(2)$ and $(3)$ in the absorption
coefficients indicates absorption by two and three nucleons respectively.\\

 

In principle we should implement other medium modifications, like the addition of a
selfenergy in the internal pion line, but the
virtual pion is far off-shell and the pion selfenergy in the denominator
of the propagator is small in comparison with the quadrimomentum of the
pion. On the other hand, we should expect some medium corrections in the
vertices such as $NN\pi$ and others but, based on calculations of \cite{Oset:1976dq}, we
expect these corrections to be very small.\\

\begin{figure}[h]
\centerline{\hbox{\psfig{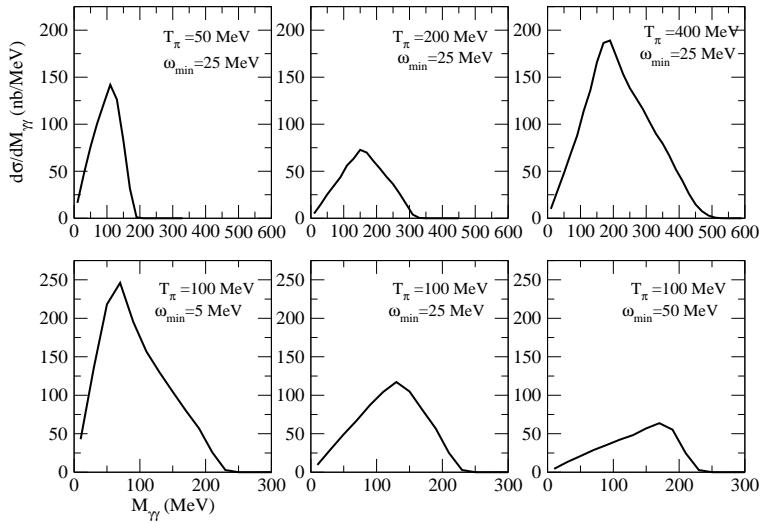}}}
\caption{\rm 
Same as Fig.~\protect{\ref{fig:z1}} but in $^{208}Pb$.}
\label{fig:z2}
\end{figure}

In Fig.~\ref{fig:z2} we present results for the invariant mass distribution
 of the two photons
for the $(\pi^-,\gamma\gamma)$ reaction in $^{208}Pb$ in the same
 cases as in Fig.~\ref{fig:z1}.
We can observe there the strong $\pi^-$ absorption
 at $T_{\pi}=200\,\textrm{MeV}$ where the initial pion is in the $\Delta$
resonance region.

\vspace{0.9cm}

\section{Looking for in medium modifications of the $\sigma$ meson}

\begin{figure}[h]
\centerline{\hbox{\psfig{file=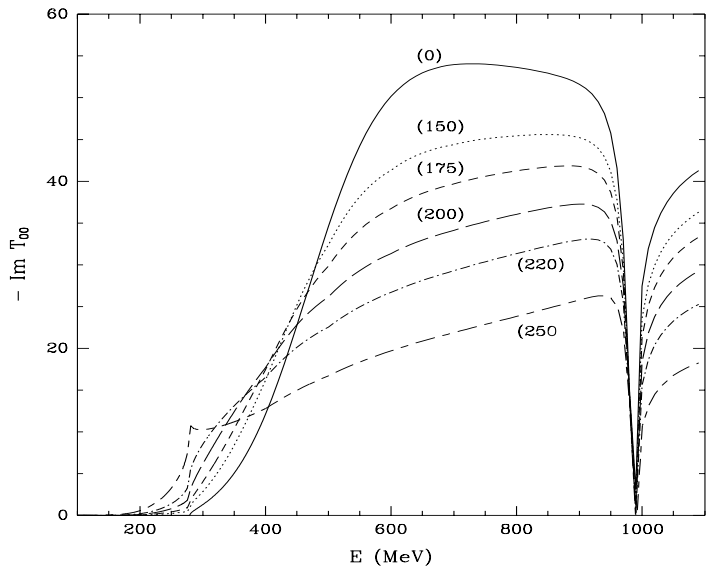,width=10cm}}}
\caption{\rm 
Imaginary part of the $\pi\pi\to\pi\pi$ in J=I=0 in the nuclear medium
for different values of $k_F$ versus the CM energy
of the pion pair.
The labels correspond to the values of $k_F$ in MeV.}
\label{fig:ImTpipi}
\end{figure}

In the work of \cite{Roca:2002vd} the authors observed a shift in the two
pion invariant mass distribution in the reaction $(\gamma,\pi^0\pi^0)$ in nuclei due to the
shift of the pole position of the $\pi\pi\to\pi\pi$ amplitude
in the scalar-isoscalar channel in
nuclear matter. When contrasted with the experimental data
of \cite{Metag}, this work represented
the first clear manifestation of the dropping of the $\sigma$ meson mass in nuclear matter.
In \cite{Roca:2002vd} the model of \cite{Chiang:1998di} for the $\pi\pi$
interaction in the nuclear medium was used. In \cite{Chiang:1998di}
the $\pi\pi$ rescattering in nuclear matter was
done renormalizing the pion propagators in the medium
and introducing vertex corrections for consistency. The results obtained with the model of
\cite{Chiang:1998di} for the imaginary part of the  $\pi\pi\to\pi\pi$ in $I=0$ as a function of the
nuclear density can be seen in Fig.~\ref{fig:ImTpipi}. 

One can see in Fig.~\ref{fig:ImTpipi} that, as the nuclear density
increases, there is a shift of strength to low $\pi\pi$ masses. This was the effect
shown in \cite{Roca:2002vd} and \cite{Metag}.

In view of the results of these works one may wonder if the $(\pi^-,\gamma\gamma)$ on nuclei is also a
suitable reaction to test this dropping of the $\sigma$ mass in nuclear matter,
because the $\pi\pi\to\gamma\gamma$ reaction is a very important mechanism in this
reaction, see Fig~\ref{fig:diagrams_free},
 and the $\sigma$ channel is a relevant part in the $\gamma\gamma\to\pi^+\pi^-$
amplitude as shown in \cite{Oller:1997yg}.\\

\begin{figure}[h]
\centerline{\hbox{\psfig{file=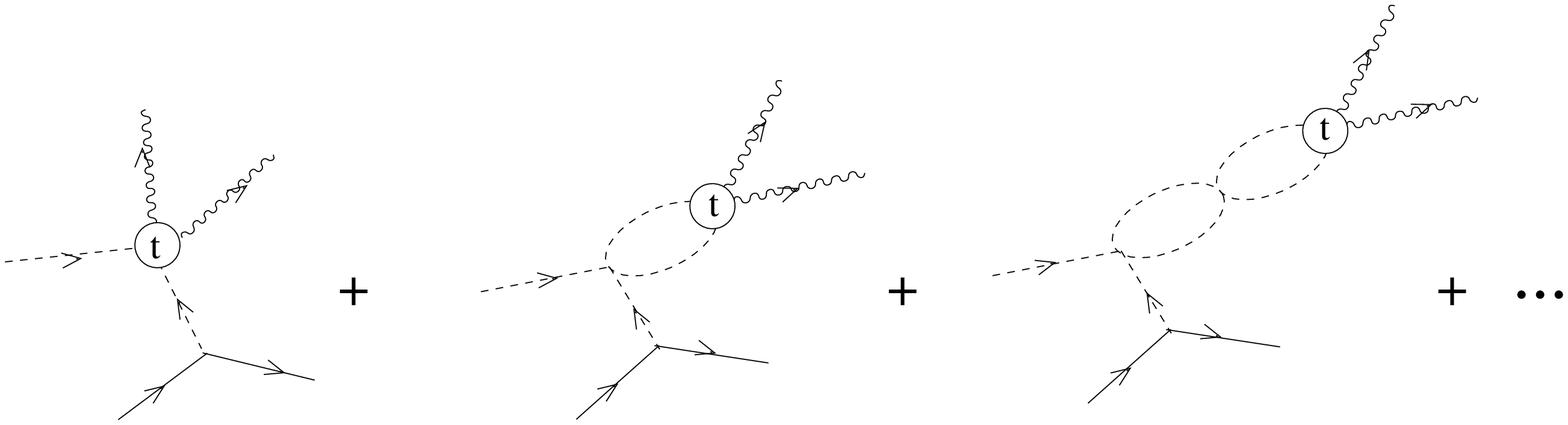,width=15cm}}}
\caption{\rm 
Rescattering of the pions for the dynamical generation of the $\sigma$ meson in the
$(\pi^-,\gamma\gamma)$ reaction.}
\label{fig:rescatt}
\end{figure}

In order to answer this question we have to implement the rescattering of the pions
in (I=0,L=0) ($\sigma$ channel) in the diagrams where they lead to two photons,
in the way of Fig.~\ref{fig:rescatt}.
Then we have to compare the invariant mass distribution of the two photons
when the $(\pi^-,\gamma\gamma)$ reaction is done in nuclei with
 that where the reaction is done on
hydrogen, and then see if there is a move of strength,
 manifesting the effect shown close to $2m_{\pi}$ in Fig.~\ref{fig:ImTpipi}.
 
\begin{figure}[h]
\centerline{\hbox{\psfig{file=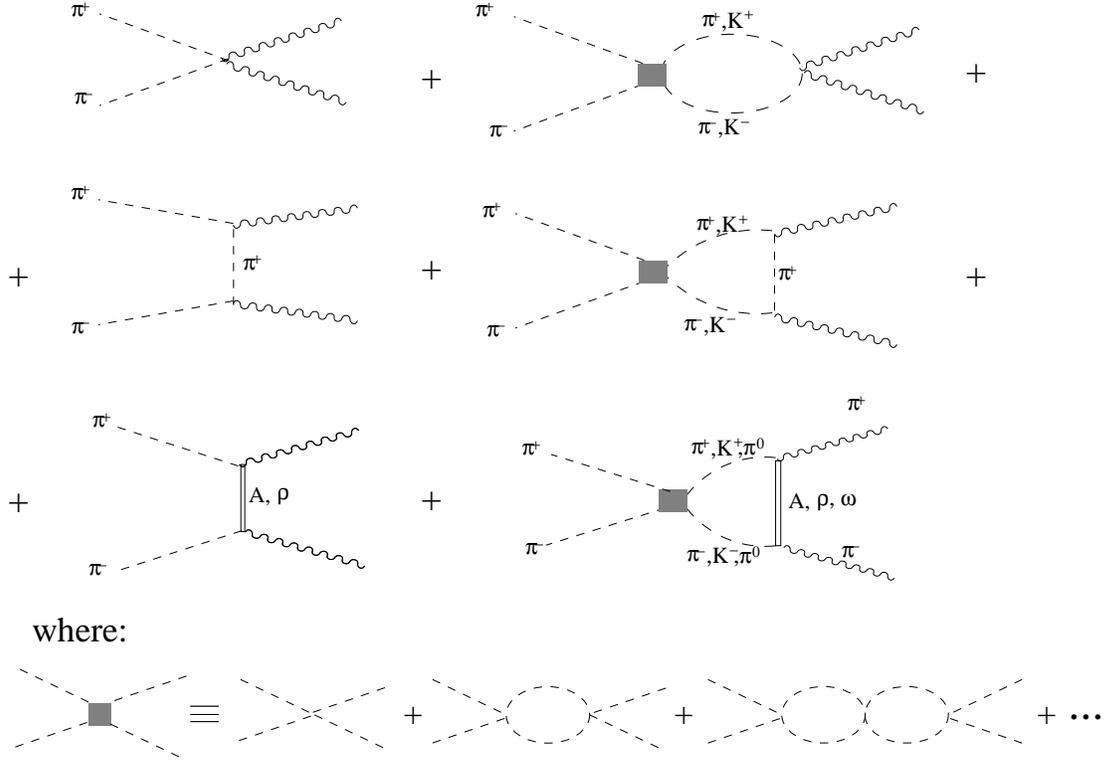,width=15cm}}}
\caption{\rm 
Diagrams for $\pi^+\pi^-\to\gamma\gamma$ included in the model of \cite{Oller:1997yg}.}
\label{fig:loops}
\end{figure}

For the $\pi^+\pi^-\to\gamma\gamma$ reaction in free space we will use the model of
 \cite{Oller:1997yg}. In this work the authors considered the resummation of the
Born terms, those obtained from chiral
Lagrangians up to order $\mathcal{O}(p^4)$ and the terms
with exchange of one axial or vector meson. The diagrammatic representation
is shown in Fig.~\ref{fig:loops} (see Ref.~\cite{Oller:1997yg} for details and meaning of
these terms).
 
The amplitude for the $\pi^+\pi^-\to\gamma\gamma$ reaction corresponding to the diagrams of 
Fig.~\ref{fig:loops} reads

\begin{eqnarray} \nonumber
t &=& t_{Born} + t_A + t_\rho + (\tilde{t}_{A \pi^+ \pi^-}
+ \tilde{t}_{\rho \pi^+ \pi^-} + \tilde{t}_{\chi \pi}) t_{\pi^+ \pi^- ,
\pi^+ \pi^- }\, + \\
&+& (\tilde{t}_{A K^+ K^-} + \tilde{t}_{\chi K}) t_{K^+ K^-,
\pi^+ \pi^-} + (\tilde{t}_{\rho \pi^0 \pi^0}+
\tilde{t}_{\omega \pi^0 \pi^0}) t_{\pi^0 \pi^0, \pi^+
\pi^-}
\label{eq:tc}
\end{eqnarray}

\noindent
where
\begin{eqnarray}\nonumber
t_{\pi^+ \pi^- , \pi^+ \pi^-} &=& \frac{1}{3} t^{I = 0} + \frac{1}{6}
t^{I = 2}\\ \label{eq:MM}
t_{\pi^0 \pi^0 , \pi^+ \pi^-} &=& \frac{1}{3} t^{I = 0} - \frac{1}{3}
t^{I = 2}\\ \nonumber
t_{K^+  K^- , \pi^+ \pi^-} &=& \frac{1}{\sqrt{6}} t^{I = 0}
\end{eqnarray}
The analytical expressions for $t_{A,\rho}$ and $\tilde{t}_i$ can be found in
\cite{Oller:1997yg}.
The meson scattering amplitudes of Eqs.~(\ref{eq:MM}) were evaluated in \cite{Oller:1997ti} by
solving the Bethe-Salpeter equation in coupled channels with the kernels formed from the lowest
order meson-meson chiral Lagrangian amplitude. This is the way the
$\sigma$ meson is  dynamically generated. 
From now on we will call, for brief, "chiral terms"
those of Eq.~(\ref{eq:tc}) without
$t_{Born}$.

When we go to the nuclear medium we have to replace the free $\pi\pi\to\pi\pi$ in $I=0$ and the
two pion loop function, $G_{\pi\pi}$, by their
corresponding values in nuclear matter, obtained in \cite{Chiang:1998di}, calculated at the local
density corresponding to the integration point in Eq.~(\ref{eq:X-IM}). 
When replacing the free $G_{\pi\pi}$ by its in-medium expression 
one automatically takes into account
the terms involving loops in the third row 
in Fig.~\ref{fig:loops}, because in 
\cite{Oller:1997yg} it was shown that these diagrams can be expressed in terms of $G_{\pi\pi}$,
factorizing on-shell the rest of the amplitude. For the loops implicit in $\tilde{t}_{\chi}$,
the loops in the first two rows of Fig.~\ref{fig:loops}, we
have just multiplied the expression given in \cite{Oller:1997yg}
by $G_{\pi\pi}(\rho)/G_{\pi\pi}(\rho=0)$.

With all these considerations we can pass now to study
if these modifications have appreciable consequences when they are
introduced in the full $(\pi^-,\gamma\gamma)$ model in nuclei.
 In order to look for the effect shown in
Fig.~\ref{fig:ImTpipi} we have to choose a kinematical region where this effect is maximized. We have
chosen a kinetic energy of the pion of $380\,MeV$ because the phase space
distribution of the two
photons makes more relevant the region of 200 to 400 MeV where the changes of the $\pi\pi$
amplitude show up. The cut in the energy of the photons has been chosen as 
$50\,$MeV in what follows in order to reduce the strength far away from the region of
interest ($2m_{\pi}$).

\begin{figure}[h]
\centerline{\hbox{\psfig{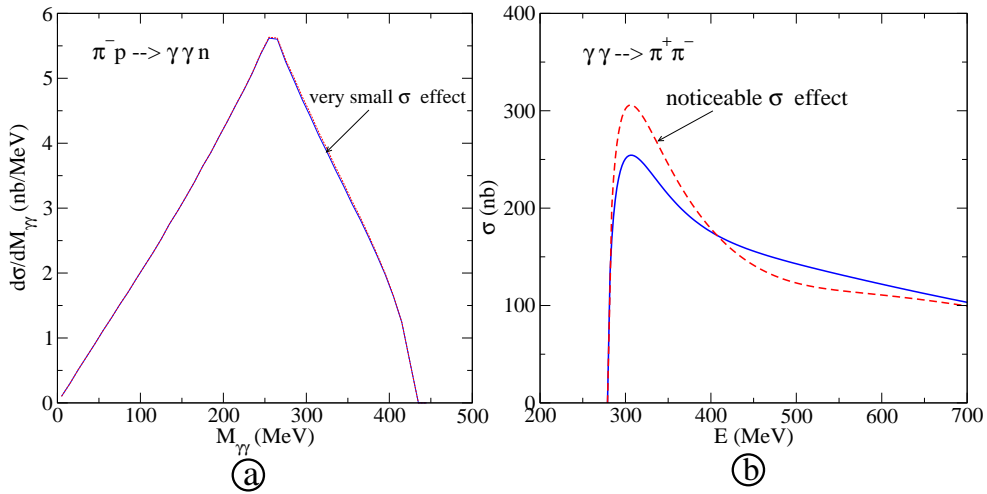}}}
\caption{\rm 
Invariant mass distribution and total cross section for $\pi^-p\to\gamma\gamma n$
and $\gamma\gamma\to\pi^+\pi^-$ reaction respectively. Continuous line: full model
without the inclusion of the chiral terms. Dashed line: full model + chiral terms. }
\label{fig:z3}
\end{figure}

\begin{figure}[h]
\centerline{\hbox{\psfig{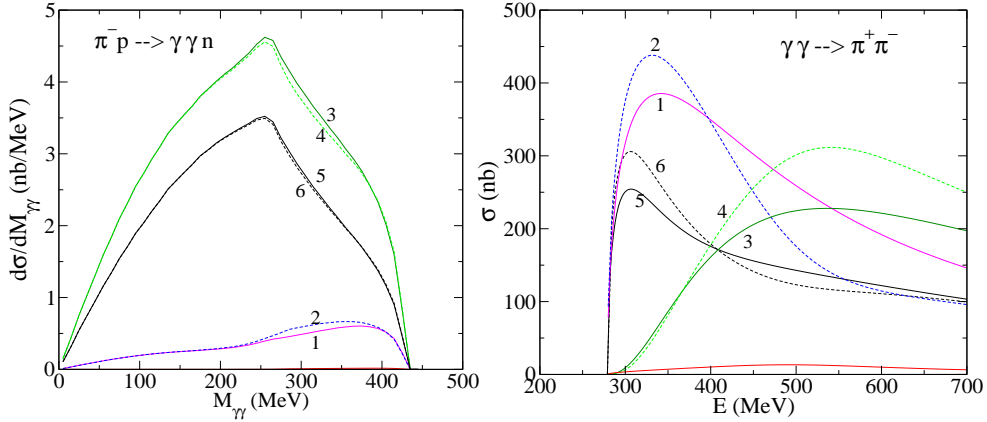}}}
\caption{\rm 
Effect of the chiral terms on the different contributions to the
$(\pi^-,\gamma\gamma)$ and $\gamma\gamma\to\pi^+\pi^-$ reactions. 
See text for explanation.}
\label{fig:z4}
\end{figure}

In Fig.~\ref{fig:z3}.a we can see the invariant mass distribution of the two photons for the
$\pi^-p\to\gamma\gamma n$ reaction with and without the explicit inclusion of the 
chiral terms, dashed and continuous lines respectively, but we
can see that the effect is hardly visible, around
1-2\%.
In Fig.~\ref{fig:z3}.b we have plotted the equivalent curves
to Fig.~\ref{fig:z3}.a but
for $\gamma\gamma\to\pi^+\pi^-$.
Looking at Fig.~\ref{fig:z3}.b one can see that the effect of the
chiral terms, where the $\sigma$ meson plays a role, is to increase the cross section in
around $20\,\%$ at $\sim 320\,MeV$ of energy. 
Therefore one could expect, in principle, a similar effect in the
$(\pi^-,\gamma\gamma)$ reaction.
But in the $(\pi^-,\gamma\gamma)$ reaction we
only obtain a tiny effect. This is due to a subtle combination of interferences
and the allowed phase space. In order to clarify this interference we show
in Fig.~\ref{fig:z4} different contributions separately, and we refer
to this figure in what follow.

 We have checked that the effect of adding the chiral terms to the
diagram $a$ of Fig.~\ref{fig:diagrams_free} is to increase the cross section in around $20\,$\%
at $320\,MeV$, both in $\gamma\gamma\to\pi^+\pi^-$
 and in the invariant mass distribution of the
two photons in the $(\pi^-,\gamma\gamma)$ reaction
(lines 2).
 In $\gamma\gamma\to\pi^+\pi^-$ the $a$ term (line 1) is
the only significative one at this energy since $b$ and $c$ (line 3)
tend to 0 at threshold. Thus the
total effect of the chiral terms in the cross section
is a $20\,$\% close to $2m_{\pi}$ (change between line 5 and 6).
But in $(\pi^-,\gamma\gamma)$ the $b$ and $c$ terms (as well as $d$ and $e$, not present in
$\gamma\gamma\to\pi^+\pi^-$) no longer vanish (line 3) because the virtual pion momenta are
not zero.
Actually it is $7$ times larger
than the $a$ term (line 1) at $320\,$MeV. When we add the chiral terms, the
interference (line 4) is
opposite to the one of $a$ and similar in size, compensating therefore
the effect (line 6) when we add the chiral terms to the $a$, $b$ and $c$
terms together (line 5). This is the
reason why the effect of the chiral terms, accounting for the dynamical
generation of the $\sigma$, is so small in the $(\pi^-,\gamma\gamma)$ reaction. In summary: the
large size of the pion-Bremsstrahlung terms and its opposite interference with the chiral ones
mask the effect of the $\sigma$ meson. 

In spite of this discouraging result, we can try to magnify the effect of the $\sigma$ meson by
making some kind of filters or kinematical cuts in order to remove the hindering terms
$b$, $c$, $d$ and $e$ of Fig.~\ref{fig:diagrams_free}. If we turn off numerically these
terms the effects that we would expect should be those shown in the right plot of 
Fig.~\ref{fig:z5}:

\begin{figure}[h]
\centerline{\hbox{\psfig{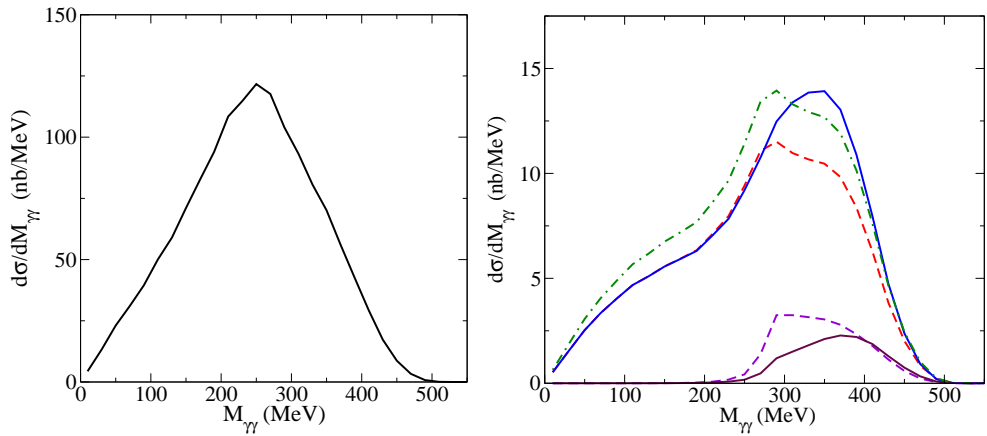}}}
\caption{\rm 
Left panel: invariant mass distribution for
$(\pi^-,\gamma\gamma)$ reaction in $^{208}Pb$ with the full
model. Right panel: without the inclusion of the mechanisms with infrared
divergence. See the text for a whole explanation.}
\label{fig:z5}
\end{figure}

In the right panel of Fig.~\ref{fig:z5} we have plotted the invariant mass distribution
 of the two photons for the
reaction in $^{208}Pb$, setting explicitly to zero the $b$, $c$, $d$ and $e$ terms
of Fig.~\ref{fig:diagrams_free} (those that manifest the infrared divergence). The plot has been
evaluated at $T_{\pi}=380\,\textrm{MeV}$ and $\omega_{min}=50\,\textrm{MeV}$.
We can observe how much strength do these contributions represent comparing to the
full model at the same energy and photon energy cut shown in the left panel.
The solid line of the right plot
represents the model with only the $a$ term plus the chiral terms
but these latter ones calculated at
zero density, that is if we turn off the density in the rescattering of the pions in the
scalar-isoscalar channel. If we turn on the density 
dependence of these rescattering terms we obtain the dashed line.
The dashed-dotted line is the same plot but renormalized to the
peak of the continuous line in order to compare both curves. A shift of
strength to low invariant masses when the medium effects are considered can be observed.
The two lower curves
represent the case when only the chiral terms are considered
(diagrams of Fig.~\ref{fig:loops} removing the first two diagrams of column one)
both at zero and
at true density
 (both magnified $7$ times in order to have the curves within the scale of the figure).
But Fig.~\ref{fig:z5} is only a theoretical exercise that shows 
the best effect we can expect,
provided one could eliminate experimentally
the $b$, $c$, $d$ and $e$ mechanisms, the last four terms
in Eq.~(\ref{eq:amplitude_Born}) respectively.
Looking at Eq.~(\ref{eq:amplitude_Born}) we can see that,
should we go to very low pion momentum, these terms would vanish, but we would be in a
kinematical region very far from $2m_{\pi}$ where the shift in the $\pi\pi$ amplitude 
in the medium occurs
according to Fig.~\ref{fig:ImTpipi}. There are, however,
some possible experimental set ups
to remove these hindering terms:

 The first one is to keep only  those events where the two photons are produced
in the direction of the incident $\pi^-$. This would force the polarization
vectors $\vec{\epsilon_1}$ and $\vec{\epsilon_2}$ to be orthogonal to $\vec{q}$
and then these terms would vanish. The photons so chosen can go in the same direction or
back-to-back. We have also to remove those which go in the same direction because they have zero
invariant mass.

\begin{figure}[h]
\centerline{\hbox{\psfig{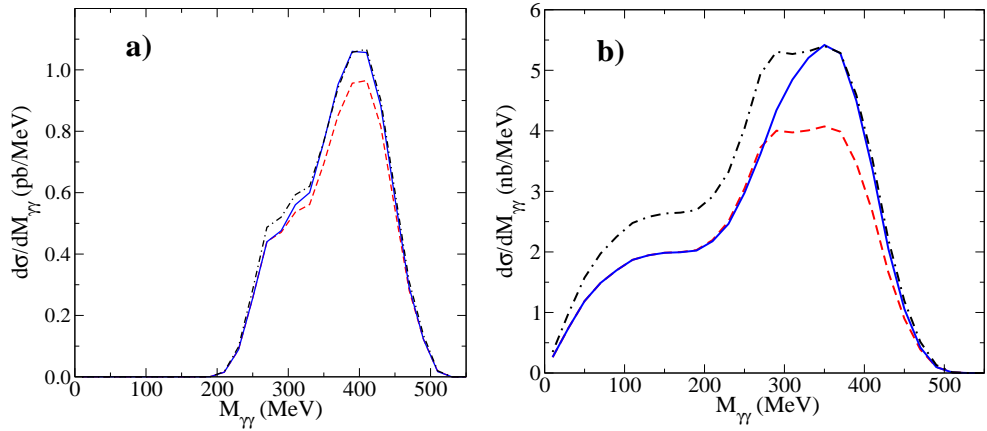}}}
\caption{\rm 
Results after implementing the filters. Left panel: only accepted events when the
photons go back-to-back and parallel to direction of the incident pion, whith an
acceptance angle of $5$~degrees. Right
panel: only accepted events when the polarization of both photons are orthogonal
to the direction of the incident pion.}
\label{fig:z6}
\end{figure}

When this filter is actually implemented, one has to choose a finite acceptance
angle for the deviation of the photons with respect to the direction of the
incident pion.  The results when we implement this filter with an acceptance angle
of $5$~degrees are shown in Fig.~\ref{fig:z6}.a.

The continuous line is calculated with zero density for the chiral terms, the dashed one is the
full model and the dashed-dotted line is the full model normalized to the continuous line. We can
see that the effect of the density in the chiral terms is visible in the 
size but barely noticeable in the shape of the strength
because this cut favours the accumulation of events at high
invariant masses, thus minimizing the $2m_{\pi}$ region. The other problem with this filter is the
strong decrease of the statistics since the cross section is reduced in nearly 
five orders of magnitude, mostly due to the reduced phase space acceptance.

Another possibility to eliminate the hindering terms, according to
Eq.~(\ref{eq:amplitude_Born}),
is to implement a filter on the polarization of the photons in such a way
that only those events with photon polarization
orthogonal to the momentum of the initial $\pi^-$ are accepted.   
In Fig.~\ref{fig:z6}.b we can see that the implementation of this latter filter causes the invariant mass
distribution to look much more like that of Fig.~\ref{fig:z5}. The shift of
the shape between the continuous and dashed-dotted lines
manifests the modification due to the nuclear medium of
the $\pi\pi$ interaction in the $\sigma$ channel. But this procedure has the strong
inconvenience of having to filter the polarization of photons in the final state.

In addition to this problem, one should point out that by filtering the photon polarization
the cross section has been reduced in a factor 20, hence reducing appreciably
the statistics of a cross section already small to start with. This also has the
inconvenience that, given the fact that an eventual polarization of the photons would
necessarily have a certain uncertainty in this polarization, this uncertainty, allowing
the contribution from the pion-Bremsstrahlung terms, could drastically enhance the cross section
with respect to the one calculated in Fig.~\ref{fig:z6}.b, thus masking the $\sigma$
in medium effects. In the same direction, the Bremsstrahlung terms on the nucleon lines,
which are small compared to the dominant terms of the full model as we pointed out, would
no longer be small compared to those of the filtered amplitude, which could also make the
cross section bigger than evaluated in Fig.~\ref{fig:z6}.b, once again blurring the effects
of the $\sigma$ in the medium.

\section{Conclusions}

We  have made a theoretical analysis of the $(\pi^-,\gamma\gamma)$ reaction
on the proton and
nuclei beyond the scope of its application to pionic atoms.
After explaining the
mechanisms considered for the free reaction, we have summarized the procedure used to study the
reaction on nuclei, based on very well tested many body techniques. As an immediate application
we have studied the viability of the reaction to test the shift of the $\sigma$
meson pole in nuclear matter, using a model which provides the dynamical generation
of the $\sigma$ meson in the $\pi\pi$ rescattering in $(I=0,L=0)$ within the framework of a
unitary extension of $\chi PT$ and its modification in nuclear
matter \cite{Chiang:1998di}. We have
implemented in the  $(\pi^-,\gamma\gamma)$ reaction the model of \cite{Oller:1997yg} for
the $\pi^-\pi^+\to\gamma\gamma$ reaction 
where the dynamical generation of the $\sigma$ plays an important
role. We have investigated whether a shift of strength
in the invariant mass distribution of the two
photons appears when we pass from the free reaction to the in-medium case.
The first observation is that the
very strong background due to the pion-Bremsstrahlung mechanisms hides the desired effect.
We have
tested several experimental filters and kinematical cuts in order to reduce
the influence of these pion-Bremsstrahlung terms, but even then we have obtained 
a small signal of the desired effects
and, furthermore, these experimental set ups seem not easy to implement.
All this said, we must conclude that the $(\pi^-,\gamma\gamma)$ reaction on nuclei
is not a very suitable reaction to test the in-medium modification of
the properties of the $\sigma$
meson, in spite of testing bigger nuclear densities than the $(\gamma,\pi\pi)$ or
$(\pi,\pi\pi)$ reactions.

Leaving this negative result aside, the present work has performed a realistic model for
the $(\pi^-,\gamma\gamma)$ reaction in nuclei, extrapolating the results for the same
reaction in pionic atoms done in \cite{Gil:1994mi} to the region of pions in the continuum.
Comparison with the experiment can serve as a further test of the many body techniques used
here which have been successfully used for other reactions, providing extra confidence in
these easy to implement and, so far, quite accurate methods, which can be applied to most
nuclear reactions at intermediate energies.

\section*{Acknowledgments}
We are specially grateful to M.~J.~Vicente~Vacas for fruitful discussions and his careful
reading of the manuscript.
One of us, L.R., acknowledges support from the Consejo Superior de
Investigaciones Cient\'{\i}ficas. This work is
partly supported by DGICYT contract numbers BFM2000-1326,
and the E.U. EURIDICE network contract no. HPRN-CT-2002-00311.

\end{document}